\documentclass[twocolumn,aps,prl]{revtex4}
\usepackage{amsmath,amssymb}
\usepackage{graphicx}
\usepackage{epstopdf}
\usepackage[usenames]{color}
\usepackage[]{natbib}
\newcommand{\be}{\begin{equation}}
\newcommand{\ee}{\end{equation}}
\newcommand{\bk}{{{\bf{k}}}}

\newcommand{\bea}{\begin{eqnarray}}
\newcommand{\eea}{\end{eqnarray}}
\newcommand{\ra}{\rangle}
\newcommand{\la}{\langle}

\newcommand{\upa}{\uparrow}
\newcommand{\dna}{\downarrow}

\newcommand{\dg}{{\dagger}}
\newcommand{\pdg}{{\phantom\dagger}}

\begin{document}

\title{Double Perovskite Heterostructures: Magnetism, Chern Bands, and Chern Insulators}
\author{Ashley M. Cook$^1$}
\author{Arun Paramekanti$^{1,2}$}
\affiliation{$^1$Department of Physics, University of Toronto, Toronto, Ontario, Canada M5S 1A7}
\affiliation{$^2$Canadian Institute for Advanced Research, Toronto, Ontario, M5G 1Z8, Canada}
\begin{abstract}
Experiments demonstrating the controlled growth of
oxide heterostructures have raised the prospect of realizing topologically nontrivial states of correlated electrons in
low dimensions. Here, 
we study heterostructures consisting of $\{111\}$-bilayers of double perovskites separated by inert band insulators. In bulk, these double
perovskites
have well-defined local moments interacting with itinerant electrons leading to high temperature ferromagnetism.
Incorporating spin-orbit coupling in the two-dimensional honeycomb geometry of a \{111\}-bilayer, we find
a rich phase diagram with tunable ferromagnetic order, topological Chern bands, 
and a $C=\pm 2$ Chern insulator regime.
%An effective two-band model of Zeeman-split $j\!=\! 3/2$ states captures this nontrivial band topology.
Our results are of broad relevance to oxide materials such as Sr$_2$FeMoO$_6$, Ba$_2$FeReO$_6$, and Sr$_2$CrWO$_6$.
\end{abstract}
\maketitle

%{\bf Introduction:}
Quantum anomalous Hall (QAH) insulators or Chern insulators (${\cal CI}$s) are remarkable topological phases which exhibit a 
quantized Hall effect even in the absence of a net magnetic field \cite{Haldane1988}. Proposals for candidate materials to
realize these phases include weakly correlated systems such as doped topological insulator (TI) films \cite{Jiang_magti} or TI interfaces \cite{FZhang2013a}, 
topological crystalline insulators \cite{FZhang2013b}, metallic chiral 
magnets \cite{Ohgushi2000,Martin2008}, silicene \cite{Zhang_silicene, Wright_silicene}, and
graphene \cite{Ding_gphn, Chen_gphn, Qiao_gphn, Tse_gphn,Nandkishore_gphn}.  Recent experiments on (Bi,Sb)$_2$Te$_3$ TI
films doped with magnetic Cr atoms have reported the first observation of the QAH effect \cite{Xue_Science2013} at temperatures $T \lesssim 0.5$K, 
although issues related to bulk conduction \cite{Ando_p3} and Cr doping inhomogeneities \cite{SGCheng2014} remain to be clarified.

A parallel significant development in recent years has been the experimental breakthrough in growing transition metal oxide (TMO) heterostructures
 \cite{Hwang_Nature2004,Mannhart_ScienceReview2010,Hwang_OxideReview_NatMat2012}.
This has motivated a significant effort towards understanding the interplay of strong electron correlations, quantum confinement, and spin-orbit coupling
(SOC), in driving 
topological states of electrons in cubic perovskites ABO$_3$, in pyrochlores A$_2$B$_2$O$_7$, or 
at oxide interfaces
\cite{Fidkowski_PRB2013, Banerjee_natphys2013, Nagaosa_ncomms2013, Michaeli_PRL2012, Okamoto_PRL2013,Hu_PRB2012, Kargarian_PRB2011,
Wang_PRL2011,Cai_arxiv2013, Yang_PRB2010,Fiete_PRB2012,Fiete_PRB2013,RuChen_PRB2013,XLi_PRL2014}. 
Realizing ${\cal CI}$s in TM oxides would be particularly useful since one expects the associated energy gaps and temperature scales to observe this phenomena 
to be significantly higher. It would also set the stage for realizing exotic correlation-driven fractional ${\cal CI}$s
\cite{Sheng2011,Neupert2011,Tang2011,Regnault2011}.

The challenge in stabilizing ${\cal CI}$s in simple TMOs stems from a delicate balance of energy scales. (i) Strong electronic correlations are crucial to 
drive magnetic order of the TM ion, thus breaking time-reversal symmetry, yet correlations should not be so strong as to cause Mott localization. (ii)
SOC on the TM ion needs to be significant to convert the magnetic exchange field into an orbital magnetic field for producing a QAH effect, yet outer shell
electrons in heavy elements with strong SOC are also typically weakly correlated and nonmagnetic.

\begin{figure}[t]
\includegraphics[scale=0.22]{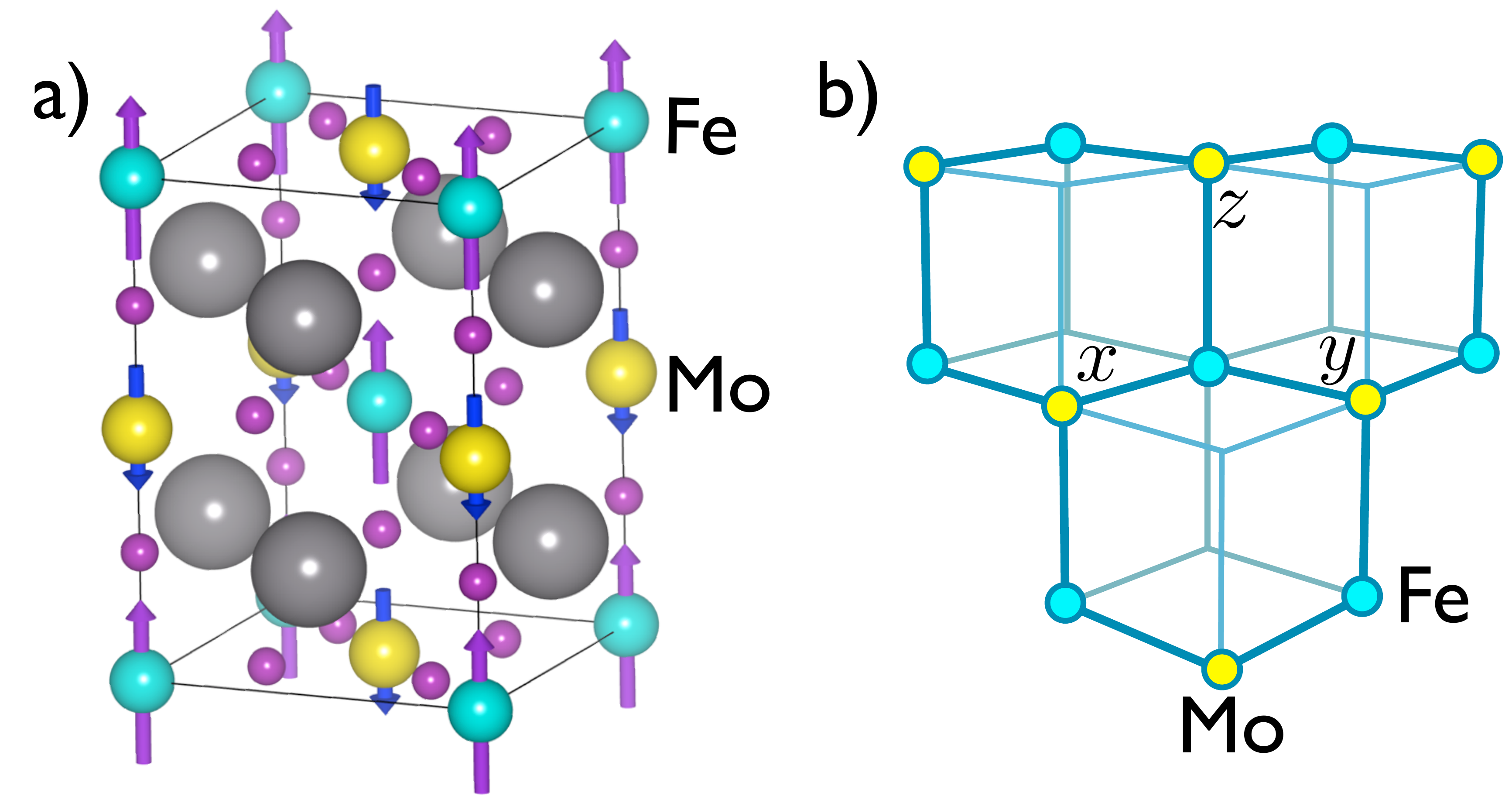}
\caption{\label{Fig:xtal} (a) Crystal structure of the double perovskite Sr$_2$FeMoO$_6$. Arrows depict ferrimagnetic
configuration of spins on the Fe and Mo sites in bulk Sr$_2$FeMoO$_6$. (b) $\{111\}$ view of a bilayer, showing buckled honeycomb
lattice with Fe and Mo ions on the two sublattices.}
\end{figure}

In this Letter,
we propose that ordered double perovskites (DPs) \cite{Serrate_Review_JPCM2007}, oxides with the chemical formula A$_2$BB'O$_6$, 
having transition metal ions B and B' residing on the two sublattices of a 3D cubic lattice as shown in Fig.~\ref{Fig:xtal}(a), can circumvent these difficulties.
For suitable choices of B, B' ions, such that B is a 3d element with strong electronic correlations driving local moment magnetism, while
B' is a 4d or 5d element which has itinerant elecrons with strong SOC, one obtains both key ingredients for realizing a ${\cal CI}$. Thus, we propose metallic 3d/4d or 3d/5d
DPs with high magnetic transition temperatures in the bulk to be promising platforms for realizing ${\cal CI}$s in a layered geometry.

We bolster this proposal by studying topological phases emerging in \{111\} bilayers of DPs sandwiched between inert band insulating 
oxides, forming a heterostructure. 
The motivation for this work stems from recent experiments on (LaNiO$_3$)$_m$-(LaMnO$_3$)$_n$ oxide superlattices grown along the $\{111\}$ direction 
\cite{Gibert_NatMat2012}  for various values of $m,n$. The $(1,1)$ superlattice, with alternately stacked triangular layers of Ni ions and Mn ions,
corresponds to the DP perovskite La$_2$NiMnO$_6$, a candidate multiferroic DP \cite{Singh_RMnNiO6_PRL2008}. Similarly, La$_2$FeCrO$_6$
has been grown artificially by alternately stacking stoichiometric LaFeO$_3$ and LaCrO$_3$ monolayers on \{111\} oriented
SrTiO3 substrate \cite{Gray_APL2010}. Here, we present our results for a Sr$_2$FeMoO$_6$ (SFMO) bilayer, as a prototypical example of a DP with high $T_c$ metallic
ferromagnetism \cite{Kobayashi1998,Sarma_PRL2000,Sarma_PRB2001,Jackeli2003,Phillips_PRB2003,Erten_PRL2011,
Vaitheeswaran_SFMO}.
Preliminary results \cite{Cook2014c}
suggest that similar physics is to be found in other materials in this family including Ba$_2$FeReO$_6$ \cite{Prellier2000,Winkler_NJP2009} 
and Sr$_2$CrWO$_6$ \cite{SCWO}.

As shown in Fig.~\ref{Fig:xtal}(b), a $\{111\}$ DP bilayer of SFMO has Fe and Mo on the two sublattices
of a (buckled) honeycomb lattice. 
The system consists of spin-orbit coupled t$_{2g}$ electrons on the triangular lattice formed by Mo, 
coupled to  local moments on the triangular Fe lattice. Our central result is the emergence, in this system, of $C=\pm 1, \pm 2$ Chern bands, 
and ${\cal CI}$s with a QAH effect, driven
by spontaneous ferromagnetism of Fe moments.

Our study of the magnetism and electronic states in the SFMO bilayer
reveals the following. Among a large variety of magnetically ordered or disordered states we have examined, 
the ferromagnetically ordered state of the Fe moments has
the lowest energy. This is consistent with experimental \cite{Kobayashi1998}  and
theoretical \cite{Erten_PRL2011} results on bulk SFMO.
The interplay of
SOC, interorbital hybridization, and a symmetry-allowed trigonal distortion leads to
different orientations of the ferromagnetic order, with distinct electronic properties.
For the $\{1\bar{1}0\}$ orientation of magnetic order, we
find electronic bands with Chern numbers $C=\pm 1$.  
For magnetic order along the $\{111\}$ direction, with Fe moments perpendicular to the bilayer, we find
that the Mo t$_{2g}$ electrons form bands with Chern numbers $C=\pm 2$;
an effective two-band triangular lattice model of  Zeeman-split $j=3/2$ states correctly captures the emergence of this 
nontrivial band topology.
These bands have a direct gap, but typically overlap in energy leading to a Chern metal.
A symmetry-allowed trigonal
distortion stabilizes a regime of a ${\cal CI}$ with $C= \pm 2$, i.e., a QAH 
insulator with a pair of chiral edge modes, having a gap $\sim 75$K.

{\bf Model:} In SFMO, strong Hund's coupling on Fe$^{3+}$ locks the 3d$^5$ electrons  into a $S_{\rm F}\!=\! 5/2$ local moment, 
which we treat as a classical spin.
% similar to Mn spins in the colossal magnetoresistive manganites \cite{Dagotto_Phys2011}.
The 4d$^1$ electron on Mo$^{5+}$ hops on or off Fe, subject to a charge-transfer 
energy $\Delta$. Pauli exclusion on Fe forces the spin of the arriving electron to be antiparallel to the underlying 
Fe moment. Kinetic energy lowering then favors ferromagnetic order of the Fe moments in bulk SFMO
\cite{Sarma_PRL2000,Sarma_PRB2001,Jackeli2003,Phillips_PRB2003,Erten_PRL2011}. Similar physics is found
in Sr$_2$CrWO$_6$ \cite{SCWO}, with a $S=3/2$ moment on Cr$^{3+}$ and an itinerant 5d$^1$ electron on W, 
as well as Ba$_2$FeReO$_6$ \cite{Prellier2000,Winkler_NJP2009} with a $S=5/2$ moment on Fe and itinerant 5d$^2$ electrons from Re. 
However, previous work has not considered the dual effect of quantum confinement and SOC in these oxides.

Here, we consider a
$\{111\}$ bilayer of SFMO, which confines electrons to a honeycomb lattice (see Fig.~1). The Mo $t_{2g}$ 
orbitals transform as $L\!=\! 1$ angular momentum states, and experience
local SOC, $-\lambda {\vec L}\cdot \vec S$, with $\lambda > 0$, leading to a low energy
$j=3/2$ quartet and a high energy $j=1/2$ doublet. Finally, the reduced symmetry of the honeycomb 
bilayer in a thin film grown along $\{111\}$ permits a trigonal distortion 
\cite{Yang_PRB2010} $H_{\rm tri}= \chi_{\rm tri} \! ({\vec L}.\hat{n})^2$, where $\hat{n}$ is 
a unit vector perpendicular to the bilayer; $\chi_{\rm tri} \! > \! 0$ corresponds to compressing the Mo oxygen octahedral cage  \cite{footnote.trigonal}.
Incorporating these new ingredients, we arrive at the model
Hamiltonian
\bea
\! H \!\!&=&\!  \!\!\! \sum_{\la ij\ra,\ell,\sigma} \!\! \left[ t^{ij}_\ell g^\pdg_\sigma(j) d^\dg_{i \ell \sigma} f^\pdg_{j\ell} + {\rm H.c.} \right] 
+ \Delta \!\! \sum_{i \ell} f^\dg_{i\ell} f^\pdg_{i\ell} + H_{\rm tri} \nonumber\\
\!\!&+\!\!& \!\!\!\!\! \sum_{\la\la ij\ra\ra,\ell,\sigma}\!\! \eta^{ij}_{\ell\ell'} d^\dg_{i \ell \sigma} d^\pdg_{j \ell' \sigma} +
i \frac{\lambda}{2} \! \sum_{i} \varepsilon^\pdg_{\ell m n} \tau^n_{\sigma\sigma'} d^\dg_{i \ell \sigma} d^\pdg_{i m \sigma'}. \label{eq:model}
\eea
Here $d$ $(f)$ denotes electrons on Mo (Fe), $i$ labels sites, $\sigma$ is the spin label, $\ell \!=\! 1,2,3$ 
($\equiv yz,zx,xy$) is the orbital index, and $\varepsilon$ is the totally antisymmetric tensor. 
With $\hat{F} = (\sin\theta\cos\phi,\sin\theta\sin\phi,\cos\theta)$ denoting the Fe moment
direction, Pauli exclusion leads to a single spin projection \cite{Erten_PRL2011} (antiparallel to $\hat{F}$) for electrons on
Fe, with $g_\upa(j)= \sin\frac{\theta_j}{2} {\rm e}^{-i\phi_j/2}$ and $g_\dna(j) = -\cos\frac{\theta_j}{2} {\rm e}^{i\phi_j/2}$.
Matrix elements $t^{ij}$ correspond to intra-orbital Mo-Fe hoppings $t_\pi$,$t_\delta$, while $\eta^{ij}$ encodes
Mo-Mo intra-orbital hopping amplitudes $t',t''$ and inter-orbital hopping amplitude $t_m$ (see Supplementary 
Material for details of hopping processes).

Such a Hamiltonian, with
strong SOC and $\chi_{\rm tri}=0$, has been shown 
\cite{Cook_PRB2013} to capture the phenomenology of the 
bulk Ba$_2$FeReO$_6$, quantitatively explaining
its band dispersion \cite{Jeon2010}, saturation magnetization \cite{Prellier2000,Teresa2007}, 
the spin and orbital polarizations \cite{Azimonte2007}, 
and spin dynamics observed using neutron scattering \cite{Plumb_PRB2013}.
For SFMO, our model captures the key energy scales: (i) the implicit strong Hund's coupling on Fe$^{3+}$ ($\sim\! \! 1$ eV, a value typical for
3d TM ions \cite{fazekas}), (ii) the Fe-Mo charge transfer energy 
($\Delta\!\! \sim \!\! 0.5$eV) \cite{Sarma_PRL2000,Erten_PRL2011}, (iii) the nearest neighbor intra-orbital Mo-Fe hopping which leads to electron 
itinerancy ($t^{ij}_\ell \! \sim \! 0.25$eV) \cite{Sarma_PRL2000,Erten_PRL2011}, and
(iv) the SOC on Mo (we set $\lambda \! \sim\!0.12$eV) is similar in magnitude to Ru
\cite{Mizokawa2001,Puetter2012}. (v) Finally, second neighbor intra-orbital and inter-orbital 
hoppings ($\eta^{ij}_{\ell\ell'} \sim 0.025$eV) are weak \cite{Sarma_PRL2000,Erten_PRL2011,Puetter2012}; nevertheless, they are important to
pin the Fe moment direction, leading to a nonzero ferromagnetic 
$T_c$ in 2D.

\begin{figure}[t]
\includegraphics[scale=0.2]{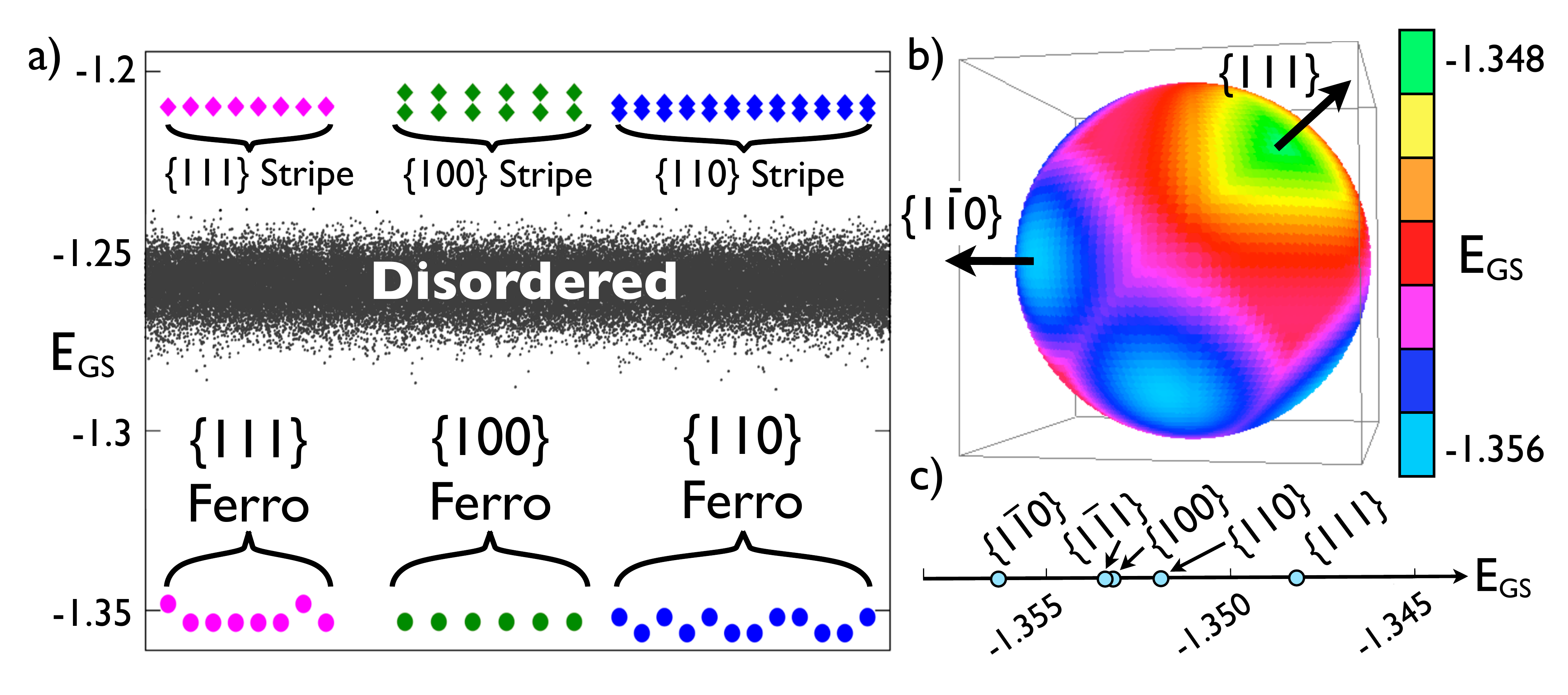}
\caption{\label{Fig:magorder} (a) Ground state electronic energy, $E_{\rm GS}$, per Fe spin in units of the Mo-Fe hopping 
$t_\pi=250$meV with $\chi_{\rm tri}=0$, shown for different Fe moment configurations
including (i) ferromagnetic, (ii) stripe-like, and (iii) disordered (random). For ordered 
states, label indicates the magnetic moment orientation. (b) $E_{\rm GS}$ for the ferromagnetic states plotted for different 
orientations of the Fe moments. (c) $E_{\rm GS}$ for the ferromagnetic states for Fe moments along high symmetry directions.}
\end{figure}

{\it\bf Magnetic ground states:} The ground state of bulk SFMO is a ferrimagnet. In order to explore the magnetic structure of the 
$\{111\}$ SFMO bilayer, we diagonalize the Hamiltonian Eq.~\ref{eq:model} with $\chi_{\rm tri}=0$, and compute the ground state
energy for various configurations of Fe moments,
including (i) ferromagnetic configurations with different spin orientations, (ii) period-2 stripe-like configurations with different spin and stripe
orientations, and (iii) random configurations.
Fig.~\ref{Fig:magorder}(a) compares these energies per Fe site, plotted in units 
of $t_\pi=250$meV which is the nearest neighbor Mo-Fe hopping amplitude, showing that the ferromagnetic states have the 
lowest energy, consistent with the kinetic energy lowering due to maximal electronic delocalization. From the energy difference between the 
ferromagnetic and disordered or stripe configurations we infer an exchange 
energy between neighboring Fe moments on the triangular lattice, $J_{\rm FF} \approx \! 1.5$meV, close to the
bulk 3D value, $\approx 3$meV, estimated from theoretical calculations \cite{Erten_PRL2011}. 
The difference stems from the different
lattice geometry and the inclusion of SOC.

\begin{figure}[t]
\includegraphics[scale=0.25]{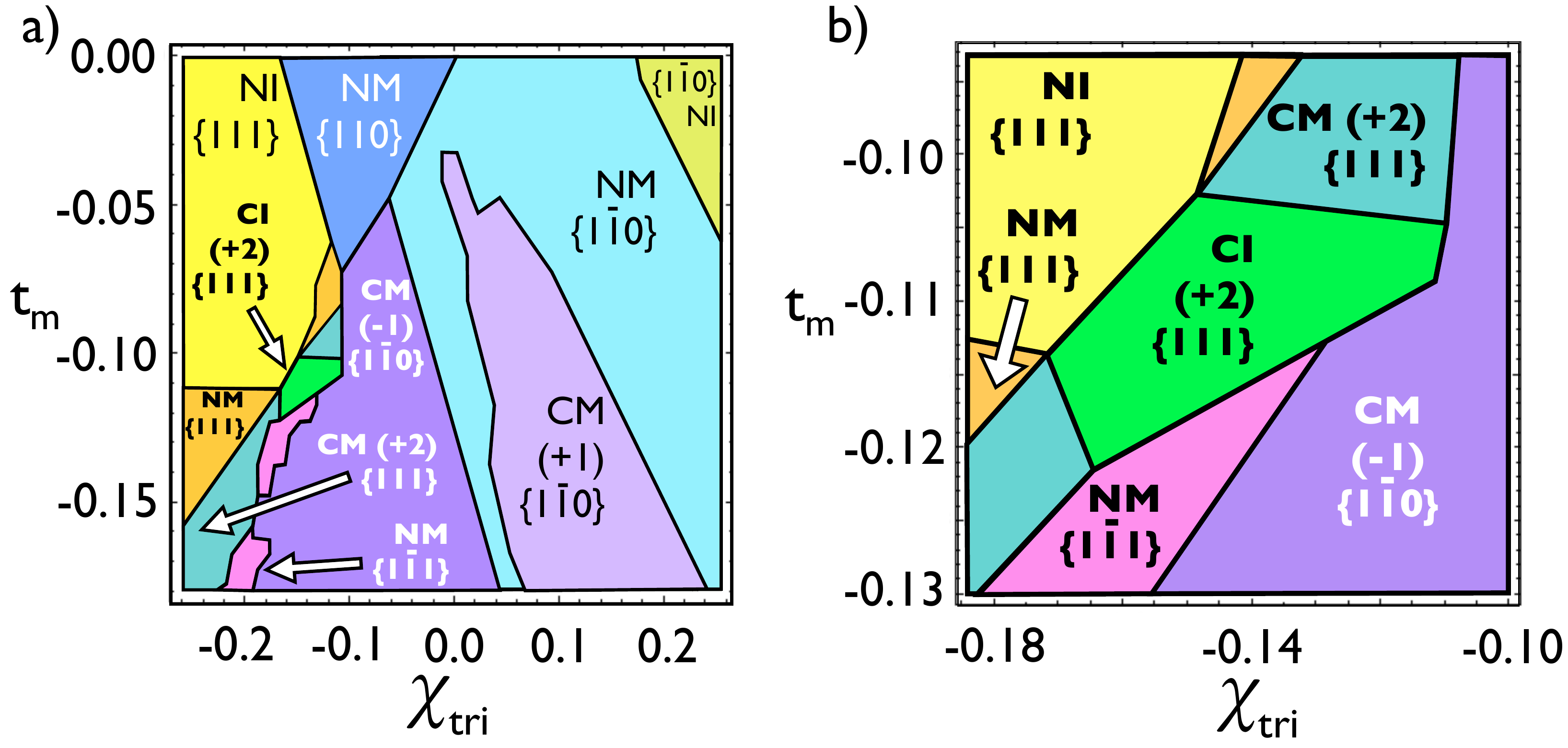}
\caption{\label{Fig:phasediag} (a) Phase diagram of the bilayer as a function of interorbital hybridization $t_m$ and trigonal distortion
$\chi_{\rm tri}$. The different phases are Chern metal (${\cal CM}$), Chern insulator (${\cal CI}$), normal metal (${\cal NM}$), and normal insulator 
(${\cal NI}$).
We have also indicated the Fe moment orientations in the different phases, and the lowest band Chern number $C$ for nontrivial band topology. 
(b) Zoomed in region showing the ${\cal CI}$ with $C=2$, and direct ${\cal NI}$-${\cal CI}$ transition.}
\end{figure}

Unlike previous work, which had Heisenberg symmetry for the magnetism \cite{Erten_PRL2011},
the inclusion of SOC leads to exchange anisotropies, resulting in 
energy differences between different ferromagnetic orientations of the Fe moments;
see Fig.~\ref{Fig:magorder}(b). With no
trigonal distortion, $\chi_{\rm tri}=0$, the six $\{1\bar{1}0\}$ orientations with Fe moments lying in the bilayer plane have the lowest energy.
As seen from Fig.~\ref{Fig:magorder}(c), other high symmetry orientations are higher in energy by $\delta E \! \sim \! 1$meV.
We have also explored the effect of trigonal distortion on the energy of different ferromagnetic orientations, keeping $\chi_{\rm tri}\neq 0$. 
For $\chi_{\rm tri} \! < \! 0$, the energy is
minimized by
$\vec L \parallel \hat{n}$. This favors the $\{111\}$ orientation of $\vec L$, and SOC then
forces the spins to also point perpendicular to the bilayer. For $\chi_{\rm tri} \! > \! 0$, it is energetically favorable to have $\vec L \! \perp \! \hat{n}$, so 
the $\{1\bar{1}0\}$ orientations remain favorable. We have numerically confirmed these expectations.
The combination of SOC and trigonal distortion thus supports a variety of ``Ising'' or ``clock''
ferromagnetic ground states. 

The broken Heisenberg symmetry induced by exchange anisotropy
leads to a nonzero magnetic $T_c$ even in the 2D bilayer. For $\{111\}$ magnetic order, with weak
anisotropy energy $\delta E$, the Ising transition temperature is implicitly given by $T_c \sim 4\pi J_{FF} S^2_F/\ln(T_c/\delta E)$
\cite{SuppMat}.
Using $J_{\rm FF} \approx \! 1.5$meV, and computed anisotropy energies across the phase 
diagram which show $\delta E \!\sim\! 0.1$-$1$ meV, we estimate $T_c \gtrsim 200$K,
lower than $T^{\rm bulk}_c \sim 400$K for bulk SFMO but still easily accessible. We next turn to the electronic properties of this 
SFMO bilayer, focusing on the band topology induced by Fe ferromagnetism.

{\bf Chern bands and phase diagram:} 
We have obtained the magnetic and electronic phase diagram of the SFMO bilayer as a function of the trigonal distortion, 
$\chi_{\rm tri}$, and the interorbital hopping $t_m$. We do this by finding the ferromagnetic orientation of the Fe atoms 
with the lowest energy, obtained by diagonalizing the Hamiltonian in Eq.~(\ref{eq:model}), and then computing the Chern number of the resulting
bands over a finely discretized Brillouin zone (BZ) \cite{Fukui2005}. 
Motivated by our finding that the magnetic order and band topology is most sensitive to $\chi_{\rm tri}$ and $t_m$, and
recent experiments showing that epitaxial strain can be used to tune the electronic structure in
TMO thin films with SOC \cite{Ramesh_Sr2IrO4film, Ramesh_SrIrO3film}, we study the ground states by
varying them over a reasonable
regime \cite{Gretarsson2013,Puetter2012}.

\begin{figure}[t]
\includegraphics[scale=0.27]{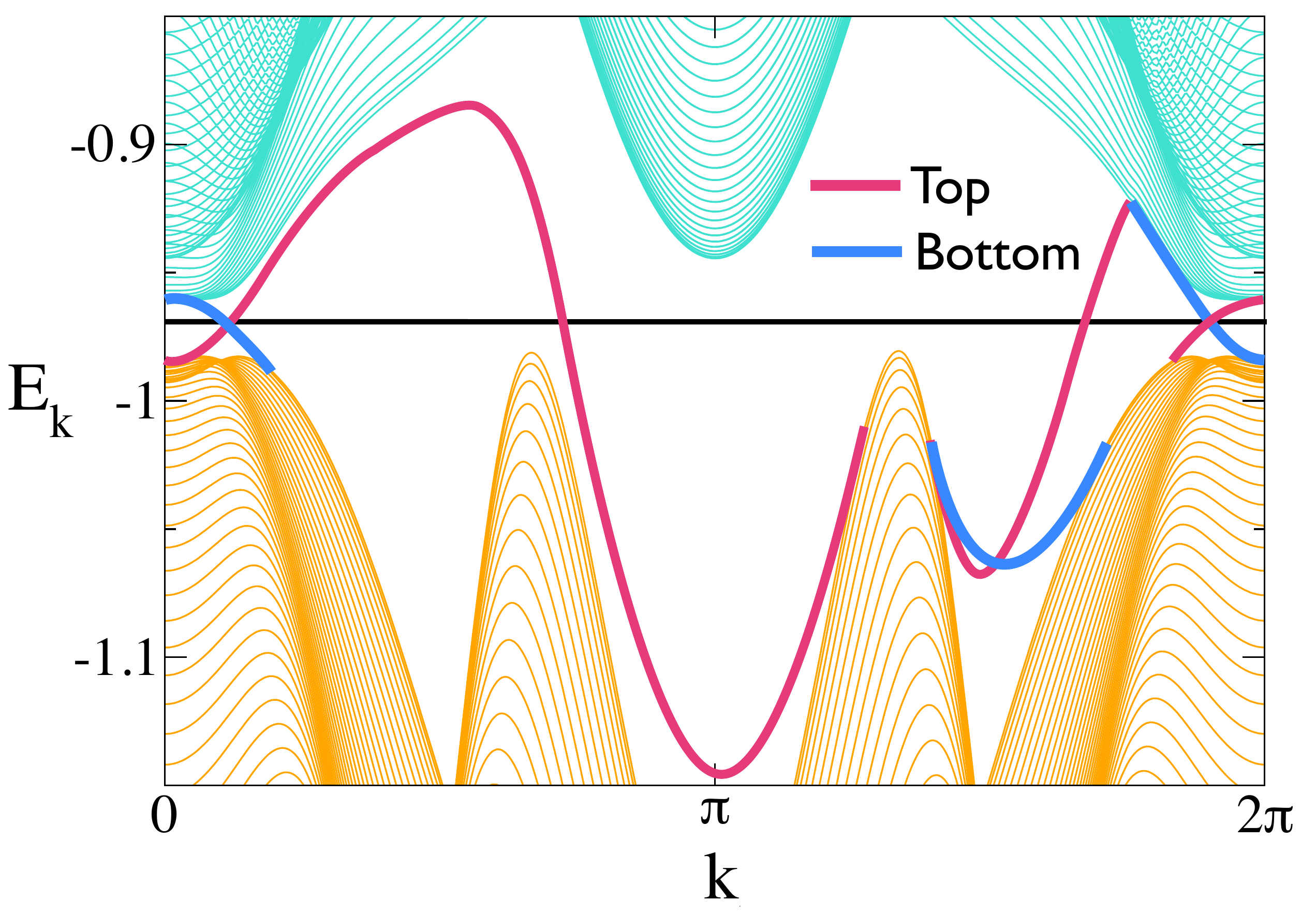}
\caption{\label{Fig:edge} Spectrum of the Chern insulator, ${\cal CI}$, in a cylinder geometry, in units of $t_{\pi}\!=\!250$meV, against momentum $k$
along the periodic direction. Here, $t_m=-0.11 t_\pi$ and $\chi_{\rm tri}=-0.15$. We find a pair of chiral
edge modes at each edge, consistent with $C\!=\!2$.
The estimated bulk gap is $0.03 t_\pi \sim 75$K.}
\end{figure}

Our calculations yield a rich phase diagram, shown in Fig.~\ref{Fig:phasediag}.
We find that the electronic states show the following phases depending on
the magnetization direction: (i) normal metal (${\cal NM}$) where the lowest pair of bands overlap in energy and they are both
topologically trivial; (ii) a normal insulator (${\cal NI}$) phase where a full gap opens up between these topologically trivial bands; (iii) A
Chern metal (${\cal CM}$) where the lowest pair of bands have nontrivial Chern numbers as indicated, yet overlap in energy, leading to
a metallic state with a non-quantized anomalous Hall response; (iv) a $C=\pm 2$
Chern insulator (${\cal CI}$) where weak trigonal distortion opens up a full gap between the two lowest topologically nontrivial bands, 
leading to a quantized anomalous Hall conductance
$\sigma_{xy}=2 e^2/\hbar$ and a pair of chiral edge modes. Fig.~\ref{Fig:edge} shows the spectrum of the ${\cal CI}$ state
in a cylinder geometry, depicting a pair of chiral modes at each edge, which cross from the valence to the conduction band. We estimate the bulk
gap of the ${\cal CI}$ state to be $0.03 t_\pi \sim 75$K.

\begin{figure}[t]
\includegraphics[scale=0.27]{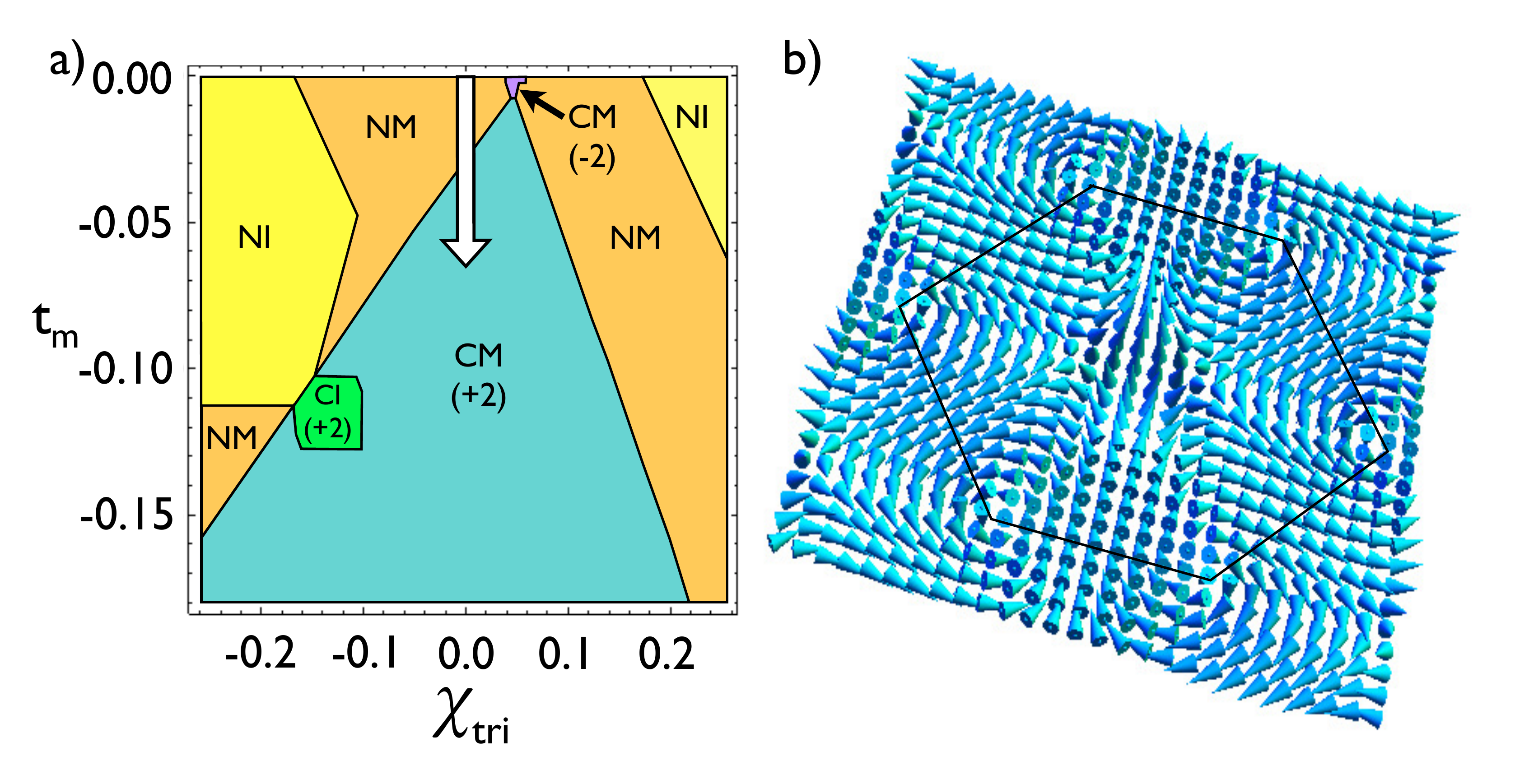}
\caption{\label{Fig:chern} (a) Phase diagram with Fe moments along $\{111\}$, showing that the ${\cal CI}$ state
arises within a wide region of $C=2$ bands. (b) Momentum space skyrmion with winding number $2$
for the ${\cal CI}$ with $C=2$.  Solid black line 
denotes hexagonal Brillouin zone.}
\end{figure}

{\bf Emergence of $C\!=\! 2$ Chern bands:} Chern bands with $C\!=\! 2$ are unusual 
\cite{FWang_Dice_PRB2011,Bergholtz_PRB2012,SYang_PRB2012,FanZhang_arxiv2013,Fang2014}
and differ from conventional Landau levels or Hofstadter bands with $C=1$. 
How can we understand the emergence of this nontrivial ${\cal CI}$? Since the $C=2$ bands
arise for magnetization perpendicular to the bilayer, we begin by studying the phase diagram with Fe moments constrained to point along 
$\{111\}$. As shown in Fig.~\ref{Fig:chern}(a), this leads to a wide swath of the phase diagram where the lowest two bands possess $C=\pm 2$.
This lowest pair of bands remains separated from the higher bands, allowing one to construct an effective two-band model to
gain insight into this physics.  

To accomplish this, we note that the predominant role of Fe moments ordered along \{111\}
is to produce an exchange field, leading to an effective Zeeman splitting of the 
spin-orbit coupled $j\!=\!3/2$ states on Mo atoms. The Chern bands arise from the lowest Zeeman split  
$j_n\!=\!-3/2,-1/2$ sublevels, where
$j_n\!=\! \vec j \cdot \hat{n}$ and $\hat{n} \parallel \{111\}$. Choosing the spin-quantization axis along $\hat{n}$, the Mo wavefunctions are:
$|j_n\!=\! -3/2\ra = \frac{1}{\sqrt{3}} (|yz \ra \!+\! \omega^2 |zx \ra \!+\! \omega |xy \ra) |\dna\ra$ and
$|j_n\!=\! -1/2\ra = -\frac{\sqrt{2}}{3} (|yz \ra \!+\! |zx \ra \!+\! |xy \ra) |\dna\ra
+ \frac{1}{3} (|yz \ra \!+\! \omega^2 |zx \ra \!+\! \omega |xy \ra) |\upa\ra$, where $\omega={\rm e}^{i2\pi/3}$. Projecting the full
model to these lowest two states (see Supplementary Material for derivation) leads to a 2-band triangular lattice
model with {\it complex} interorbital hopping. Near the $\Gamma$-point, the interorbital hopping
takes the form $\sim (k_x + i k_y)^2$; band inversion induced by increasing $t_m$ thus produces a momentum-space 
skyrmion with winding number $2$, as shown in Fig.~\ref{Fig:chern}(b), resulting in the observed $C= 2$ Chern bands. Weak trigonal 
distortion opens a full gap leading to a ${\cal CI}$.

Remarkably, the phase diagram features a direct ${\cal NI}$-${\cal CI}$ transition, as
seen from Fig.~\ref{Fig:chern}(a)). This
transition is driven by
a gap closing at the BZ center, leading to a quadratic band touching with $2\pi$ Berry phase; 
this is protected by $C_6$ lattice symmetry \cite{Sun_PRL2009}.

{\bf Discussion:}
We have shown that double perovskite metals can exhibit a variety of ferromagnetic orders
and band topologies in a \{111\} bilayer. Such Chern bands in half-metals
have also been discussed recently at CrO$_2$-TiO$_2$ interfaces \cite{Cai_arxiv2013}.  Although the topological phases we have
discussed are stable to electron interactions, such interactions are marginally relevant at the ${\cal NI}$-${\cal CI}$ quadratic band
touching transition \cite{Sun_PRL2009,Vafek_PRB2010,Zhang_PRB2012}. This leads to a window of a {\it spontaneous}
nematic ${\cal CI}$ near the ${\cal CI}$-${\cal NI}$ transition \cite{Hickey2014}. The broken inversion symmetry in the bilayer will lead to a Rashba interaction; while the
topological phases we have uncovered are stable to small Rashba coupling, a strong Rashba interaction
will drive spin spirals of Fe moments \cite{Banerjee_natphys2013,XLi_PRL2014}.
Further work is then necessary to understand
the resulting electronic phases. In future work, we will discuss bilayers of
5d-based double perovskites such as Ba$_2$FeReO$_6$ and Sr$_2$CrWO$_6$ which have a 5d$^2$ or 5d$^1$ configuration of electrons resulting in
stronger SOC, which could stabilize robust ${\cal CI}$ phases.

{\it Note added in proof.---} While this manuscript was being finalized for publication, a recent preprint has appeared also discussing 
QAH effect in \{001\} oriented double perovskite monolayers; see H. Zhang, H. Huang, K. Haule, D. Vanderbilt, arXiv:1406.4437 (unpublished).

\acknowledgments
This research was supported by NSERC of Canada. We acknowledge useful discussions with E. Bergholtz, G.A. Fiete, H.Y. Kee, S.B. Lee, S. Parameswaran,
M. Randeria, J.M. Triscone, N. Trivedi, and P. Woodward.

\appendix

\widetext

\section{Parameters in the tight binding model.}
\begin{figure}%[tbc]
\includegraphics[scale=0.2]{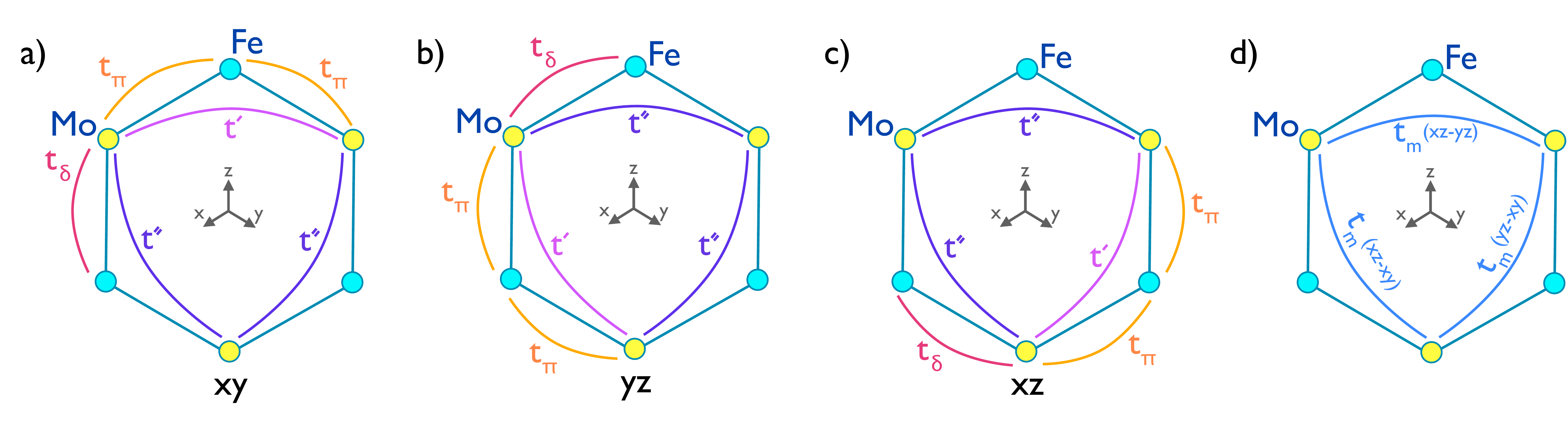}
\caption{\label{Fig:hop} Intra-orbital hopping amplitudes $t_{\pi}$, $t_{\delta}$, $t'$, $t''$ for different orbitals: (a) $xy$-orbital, (b) $yz$-orbital, (c) $xz$-orbital. 
(d) Inter-orbital hopping amplitude between pairs of indicated orbitals on Mo sites.}
\end{figure}

We consider symmetry allowed nearest neighbor Mo-Fe intra-orbital hoppings. For next neighbor Mo-Mo hoppings, intra-orbital as well as inter-orbital terms are
allowed by symmetry, and we retain both processes. The intra-orbital hopping terms are shown in Fig.~\ref{Fig:hop}(a)-(c) for $d_{xy}$,$d_{yz}$,$d_{xz}$ orbitals.
The two nearest neighbor intra-orbital hoppings are denoted by $t_{\pi}$ and $t_{\delta}$. The next-neighbor intra-orbital hoppings are denoted by
$t'$, and $t''$. Finally, Fig.~\ref{Fig:hop}(d) depicts the inter-orbital hopping, with coupling $t_m$, between different indicated orbitals on nearest pairs of Mo sites.
In our computations, with $t_\pi=1$, we set $t_\delta=-0.11$, $t'=-0.09$, $t''=0.1$, which are similar to values in the
literature \cite{Sarma_PRL2000,Sarma_PRB2001,Erten_PRL2011}. We expect a similarly small interorbital hopping $t_m \sim -0.1 t_\pi$ \cite{Puetter2012}. 
These hopping parameters
provide a good description of the bulk properties; however, they might get slightly modified due to the trigonal distortion in the
bilayer geometry.

Since the magnetic anisotropies are most sensitive to $\chi_{\rm tri}$ and $t_m$, we vary just the strength of these parameters, 
keeping $t_\delta,t'$, and $t''$ fixed. We
fix the charge transfer energy $\Delta=2.5 t_\pi$ \cite{Erten_PRL2011}, and the spin orbit coupling $\lambda=0.5 t_\pi$ as appropriate for
4d elements \cite{Mizokawa2001,Puetter2012}. We fix $t_\pi=250$meV, close to values
used in earlier studies \cite{Sarma_PRL2000,Erten_PRL2011}. We have checked that the Chern bands are robust to slight variations in these hopping parameters 
and tuning of the spin orbit coupling strength.

\section{Estimate of ferromagnetic $T_c$.}

In the absence of spin-orbit coupling, the effective model for Fe moments has full spin-rotational symmetry, leading to $T_c=0$ for ferromagnetic order in
the 2D bilayer. With spin-orbit coupling, this Heisenberg symmetry is broken to a discrete
symmetry, allowing for a nonzero $T_c$. Below, we estimate $T_c$ in the case of the Ising ordered state along $\{111\}$ which supports interesting $C=\pm 2$ Chern
bands.

We start from the isotropic 2D Heisenberg model, where the magnetic correlation length diverges as $\xi(T) \sim {\rm e}^{2\pi \rho_s/T}$ \cite{Chakravarty_PRL1988}, 
with the spin stiffness $\rho_s \sim J_{\rm FF} S^2_{\rm F}$. For weak Ising exchange anisotropy $\delta E$, the energy cost of misaligning moments away from 
the Ising axis over a correlated domain of area $\xi^2(T)$ is 
$\delta E \times \xi^2(T)$. Equating this with $T$ yields an implicit expression for the Ising ordering temperature \cite{Abanin2010}
as
\be
T_c \sim \frac{4\pi \rho_s }{\ln(T_c/\delta E)}
\ee
Using $J_{\rm FF}=1.5$meV, $S_{\rm F}=5/2$, and computed anisotropy energies $\delta E \sim 0.1$meV, yields an estimate 
$T_c \sim 250$K, which is only logarithmically sensitive 
to $\delta E$. 

Furthermore, numerical studies of Heisenberg models with weak Ising exchange anisotropy \cite{Serena_PRB1993} 
find transition temperatures which are $\sim 50\%$ of the Ising model transition temperature, even for weak anisotropies ($\sim 10^{-2}$ to $10^{-1}$). 
In our case, using this numerical result would suggest $T_c \sim 200$K, close to the above analytical estimate. This is the estimated Ising transition temperature
quoted in the paper.

\section{Effective two-orbital model of $C=\pm 2$ Chern bands}
\begin{figure}[tbc]
\includegraphics[scale=0.25]{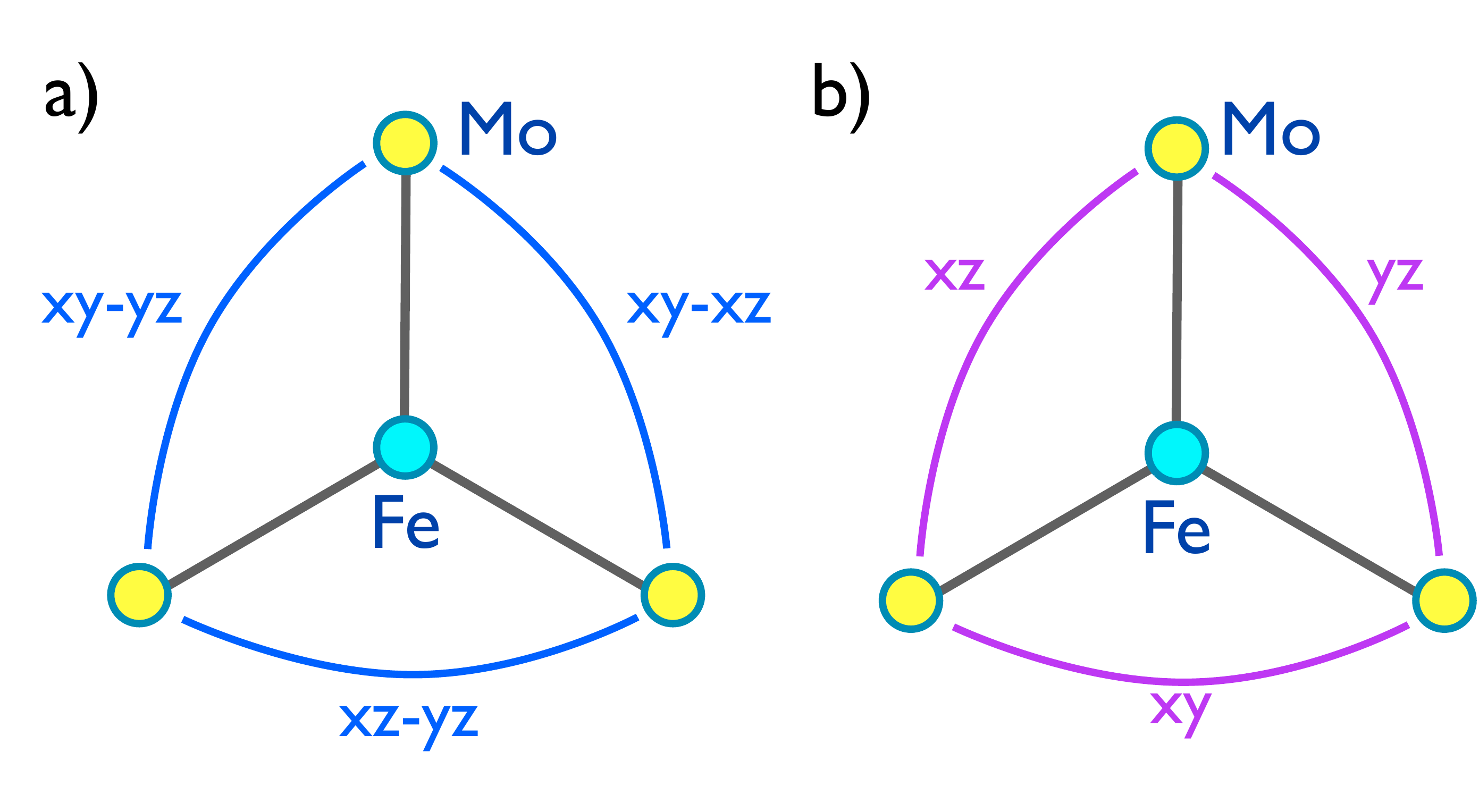}
\caption{\label{Fig:hop_2band} Hopping processes in the effective triangular lattice model of Zeeman split $j=3/2$ states on the
Mo sites. (a) Inter-orbital hopping between neighboring Mo sites. (b) Intra-orbital hopping processes between Mo sites. These
hopping processes are projected to the $j_n=3/2,1/2$ atomic states, yielding the 2-band Hamiltonian discussed above.}
\end{figure}

Here we present the derivation of the effective 2-band model which captures the formation of $C=\pm 2$ Chern bands, leading to
a simple understanding of our numerical results.
The spin-orbit coupled atomic wavefunctions corresponding to $j=3/2$ states with projection $j_n=3/2,1/2$ along the \{111\} axis 
are respectively given by
\be
|j_n\!=\! 3/2\ra = \frac{1}{\sqrt{3}} (|yz \ra \!+\! \omega |zx \ra \!+\! \omega^2 |xy \ra) |\upa\ra,
\ee and
\be
|j_n\!=\! 1/2\ra = -\frac{\sqrt{2}}{3} (|yz \ra \!+\! |zx \ra \!+\! |xy \ra) |\upa\ra
+ \frac{1}{3} (|yz \ra \!+\! \omega |zx \ra \!+\! \omega^2 |xy \ra) |\dna\ra,
\ee 
where $\omega={\rm e}^{i2\pi/3}$. Here $j_n \equiv \vec j \cdot \hat{n}$
with $\hat{n}$ along $\{111\}$, and the Fe moments are assumed to point along $\{\bar{1}\bar{1}\bar{1}\}$. Due 
to the Fe ordering, there is an effective Zeeman field experienced by the Mo sites which leads to a Zeeman
splitting $B_z$ between the $j_n=3/2$ and $j_n=1/2$ states.
Since SFMO is half-metallic, the relevant bands near the Fermi level 
are well described by considering only hopping of the $\upa$ spins, and by focusing only on the Mo sites due to 
the charge transfer energy $\Delta =2.5 t_\pi$ which suppresses occupation on Fe sites. The Mo-Mo hopping 
has two dominant contributions: (i) the
inter-orbital term $t_m$ in the original Hamiltonian; (ii) an effective $t'_{\rm eff}$ hopping, which includes the 
direct $t'$ hopping between Mo-Mo as well as indirect 
Mo-Fe-Mo hoppings which can occur at ${\cal O}(t^2_\pi/\Delta)$. These are schematically depicted in Fig.~\ref{Fig:hop_2band}.

We can project both hopping processes onto the $j_n=3/2,1/2$
atomic states, which leads to a 2-orbital triangular lattice Hamiltonian. In momentum space, this takes the form
\bea
H(\bk)= \begin{pmatrix} -\frac{2}{3} (t'_{\rm eff} - t_m) \gamma_\bk - B_z & \frac{2\sqrt{2}}{3\sqrt{3}} (t'_{\rm eff}-\omega t_m) \beta_\bk \\ 
\frac{2\sqrt{2}}{3\sqrt{3}} (t'_{\rm eff}-\omega^2 t_m) \beta^*_\bk &  -\frac{4}{9} (t'_{\rm eff} + 2 t_m) \gamma_\bk + B_z \end{pmatrix}
\label{eq:2band}
\eea
Let us define 
$\hat{a}=\hat{x}, \hat{b}=-\hat{x}/{2} + \hat{y} \sqrt{3}/2, \hat{c}=-\hat{x}/2-\hat{y} \sqrt{3}/2$. In terms of these, the matrix elements are
given by $\gamma_\bk\!=\! \sum_\delta \cos\bk\cdot\hat{\delta}$ with $\hat{\delta} \equiv \hat{a},\hat{b},\hat{c}$, and
$\beta_\bk=\omega \cos \bk\cdot\hat{a} + \omega^2 \cos \bk\cdot\hat{b}  + \cos \bk\cdot\hat{c}$. We expect $B_z \sim t'_{\rm eff}$. Fixing
$B_z,t'_{\rm eff}$ and varying $t_m$ leads to a transition between (i) a topologically
trivial state where both bands have Chern number zero and (ii) a topologically nontrivial state where bands have Chern numbers $C=\pm 2$.
This topologically nontrivial state is
characterized in momentum space by the development of a winding number $2$ skyrmion texture as shown in Fig.~5(b) of the paper,
where the arrows represent the `effective magnetic field' direction in the $2\times 2$ space of Eq.~\ref{eq:2band}.


\begin{thebibliography}{59}
\bibitem{footnote.trigonal}
For the Mo 4d$^{1}$ configuration, this trigonal
distortion could occur as a spontaneous Jahn-Teller effect.
\end{thebibliography}
\end{document}